\documentclass{osa-article}
\journal{oe}
\articletype{Research Article}
\begin{document}

\title{Multi-surface catadioptric freeform lens design for ultra-efficient off-axis road illumination}

\author{ShiLi Wei,\authormark{\#} ZhengBo Zhu,\authormark{\#}  ZiChao Fan,\authormark{} YiMing Yan\authormark{} and
DongLin Ma\authormark{*}}
 \address{\authormark{}School of Optical and Electronic Information, Huazhong University of Science and Technology, Wuhan, Hubei 430074, China}
\address{\authormark{\#}These authors contribute equally}
\email{\authormark{*}madonglin@hust.edu.cn} %% email address is required

% \homepage{http:...} %% author's URL, if desired

%%%%%%%%%%%%%%%%%%% abstract %%%%%%%%%%%%%%%%
%% [use \begin{abstract*}...\end{abstract*} if exempt from copyright]

\begin{abstract}
We propose a novel design methodology to tackle the multi-surface catadioptric freeform lens design for off-axis road illumination applications based on an ideal source. The lens configuration contains an analytic refractive entrance surface, an analytic total internal reflective (TIR) surface and two freeform exit surfaces.  A curl-free energy equipartition is established between the source and target plane and divided to implement the composite ray mapping mechanism. Furthermore, the analytic TIR surface and refractive entrance surface are optimized for the minimal Fresnel losses and surface error based on genetic algorithm (GA). The results show a significant improvement on illuminance uniformity and ultra-high transfer efficiency compared to our proposed result in [Zhu et al., Opt. Exp. 26, A54-A65 (2018)].
\end{abstract}

%%%%%%%%%%%%%%%%%%%%%%%%%%  body  %%%%%%%%%%%%%%%%%%%%%%%%%%
\section{Introduction}

 Many basic approaches are developed for the design of freeform illumination optics based on an ideal source assumption such as ray mapping method\cite{ding2008freeform,bauerle2012algorithm,bruneton2013high,desnijder2019ray,ma2015tailoring}, supporting quadric method (SQM)\cite{oliker2017controlling,oliker2006freeform,michaelis2011cartesian} and Monge-Amp\'{e}re equation (MA)\cite{wu2013freeform}. However, the current researches mainly focus on on-axis illumination design with single surface\cite{wu2018design}. Compared with on-axis illumination, off-axis illumination has more widespread applications in pratical life and is more challenging to achieve. Formulating appropriate freeform secondary optics for off-axis road illumination is still a tough issue due to the high requirements of illuminance uniformity and efficiency. Because of the total internal reflection (TIR) caused by the large deflection angle on the negative off-axis direction, it is generally impossible to achieve the design objectives with single freeform exit surface and a spherical entrance surface. Exploring efficient methodologies for the design of multiple surfaces and creating novel configurations for ultra-efficient off-axis illumination systems are still urgent and rewarding.
 
 One of the most famous multiple freeform surface design method is SMS3D proposed by Pablo Ben\'{i}tez and Juan C.Mi$\widetilde{n}$ano\cite{gimenez2004simultaneous}. It allows the generation of N optical freeform surfaces coupling N pairs of ray bundles. The SMS3D method is a robust and efficient algorithm for extended sources, which is beyond the discussion scope of our research. 
 
 In paper\cite{bauerle2012algorithm,bruneton2013high}, the authors presented a two step method for double freeform surfaces design. The solution is implemented by establishing a ray mapping between the source and target plane and optimizing double freeform surfaces for the light flux difference, smoothing component and distance to target's boundary\cite{bruneton2013high}.  Unfortunately, the statement of detailed numerical process is too simple for other researchers to duplicate this work.
 
Mikhail A.Moiseev et al. have done many beneficial researches for the design of multiple freeform surfaces\cite{moiseev2014design,moiseev2015design,kravchenko2017development,moiseev2011design}. In paper\cite{moiseev2014design}, the authors presented a novel optimization method for designing double freeform surfaces specified by scalar bicubic spline function. In paper\cite{moiseev2015design,kravchenko2017development}, the authors developed
a modification of supporting quadric method (SQM) for computation of the optical element with two smooth surfaces for accurate generation of required prescribed intensity distribution.

Wu et al. proposed a general mathematical model of two freeform surfaces lens design for point-like source\cite{wu2018formulating}. The two freeform surfaces design can be converted to an elliptical Monge-Amp\'{e}re equation with a predefined entrance surface and yield a better solution to prescribed illumination design problem.

As the methods listed above, most of them are exploring the design for on-axis freeform illumination system. There is still lack of superior freeform optics solutions for off-axis illumination applications. In this paper, we propose a novel multi-surface freeform catadioptric lens configuraion with four elaborately designed surfaces as well as an efficient methodology combined with the ray mapping method and optimization process based on genetic algorithm (GA). The results show superior optical performance and ultra-high collection efficiency compared to our proposed result in Ref.\cite{zhu2018catadioptric} for off-axis road illumination.

\section{Statement of the problem}

The cross-section profile of the conceptual design configuration is presented in Fig. 1. The lens configuration contains an analytic refractive entrance surface, an analytic total internal reflective (TIR) surface and two freeform exit surfaces. The rays on the positive off-axis direction  are collected and redirected by an analytic entrance surface, then regulated by a freeform exit surface to achieve the prescribed energy distribution on target plane. The rays on the negative off-axis direction are collected by a flat surface, redirected by an analytic TIR surface and the second freeform exit surface.

\begin{figure}[h!]
\centering\includegraphics[]{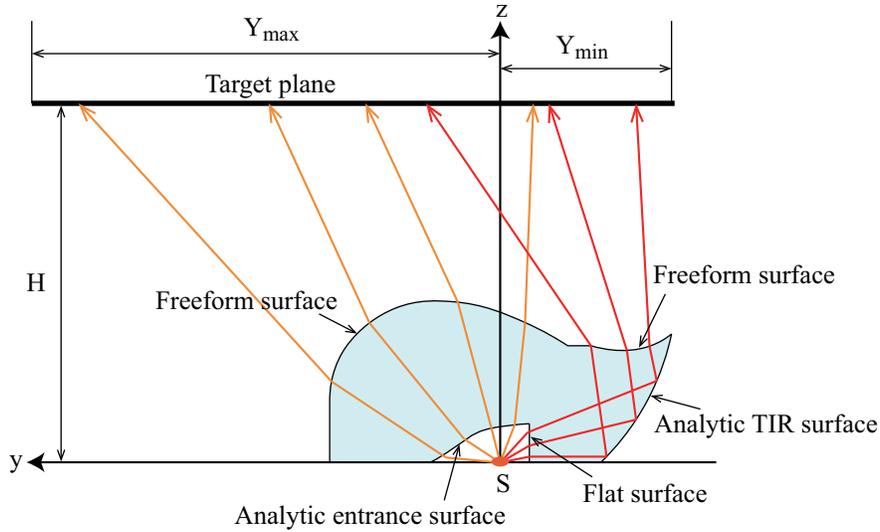}
\caption{The conceptual design configuration of multi-surface catadioptric freeform lens.}
\end{figure}

The configuration of the catadioptric freedom lens provides adequate design freedom and is able to control the stray light on the negative off-axis direction, which shows the potential of the configuration to provide a superior solution for off-axis illumination.

\section{Method}

In the previous publication \cite{zhu2018catadioptric}, the authors employed the composite ray mapping method\cite{ma2015freeform} which combined the "double pole" ray mapping technique\cite{ma2015tailoring} and the $\theta$-$\phi$ ray mapping technique to establish a diffeomorphism so that the transformed irradiance distribution matches the target distribution. Furthermore, the Euler's iteration algorithm is implemented to construct the freeform exit surface and the freeform TIR surface in\cite{zhu2018catadioptric}. In this paper, we completely updated the design method and lens configuration for preferable optical performance. The design method will be demonstrated detailedly in this section. 

\subsection{Establish a curl-free energy equipartition correspondence}

The ray mapping method is generally an efficient and flexible approach to freeform illumination design. It's not being well addressed to find a mapping that satisfies the surface integrable condition until Karel Desnijder et al. presented a new mapping scheme that alter an initial mapping via a sympletic flow of an equi-flux parametric coordinate system\cite{desnijder2019ray}. However, the integrable ray mapping method for multiple freeform surfaces design is still unsolved especially for the case that the entrance surface is changed. Second best, the optimal transport (OT) map is always used to achieve the approximated solution for freeform illumination. Although the surface normal fields are integrable into continuous surface only in the paraxial regime\cite{bosel2016ray}, the OT mapping has been proven to be able to yield a favorable design for multi-surface freeform illumination design in many applications\cite{bauerle2012algorithm,bruneton2013high,bosel2016ray,feng2013beam,feng2013designing}.  In this section, we will employ the optimal transport mapping method to establish a ray correspondence between source and target plane.

We use the Cartesian coordinate system to denote the source and target plane. For large emission angle source such as LED, it is universal to describe the light distribution using intensity defined as a position function on the whole unit hemisphere. We project the light distribution of this hemisphere onto a unit circle on the plane of the light source as shown in Fig. 1(a).  Easy to prove, the equal irradiance on the source plane from a Lambertian emitter whose intensity distribution is $I_{0}cos\theta$ ($\theta$ is the polar angle and $I_{0}$ is the peak luminous intensity on the normal direction of light source) is uniform. Assume that $\textbf{x}=(x_{1},x_{2})\in\mathcal{X}$ denotes the coordinate and boundary of source plane, $\textbf{y}=(y_{1},y_{2})\in\mathcal{Y}$ denotes the coordinate and boundary of target plane, $E_{s}(x_{1},x_{2})$ and $E_{t}(y_{1},y_{2})$ represent the irradiance distribution of source and target plane.

\begin{figure}[h!]
\centering\includegraphics[width = 13cm]{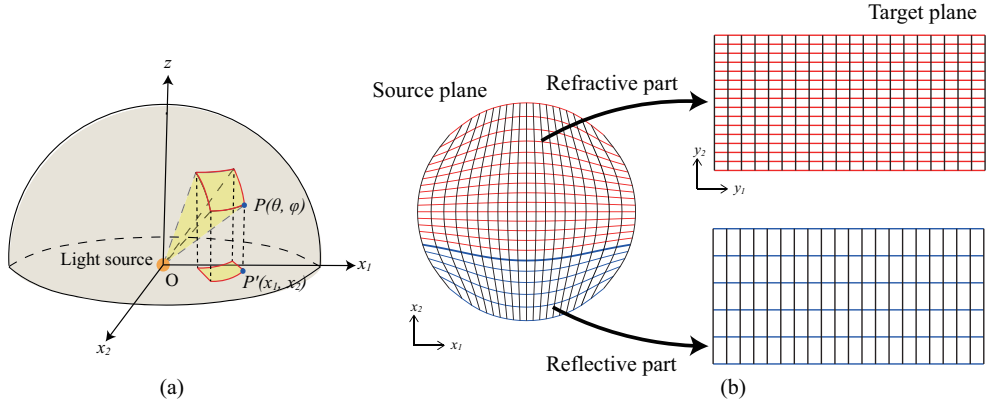}
\caption{(a)Cartesian coordinate system for source plane, (b)curl-free mapping between source and image plane.}
\end{figure}

The optimal transport mapping is the unique gradient of a convex function, which means the mapping is curl-free. Let $\bigtriangledown$$u$ labels this gradient. Substituting $\textbf{y}$=$\bigtriangledown$$u$ to the energy conversation equation, the problem can be converted to 

\begin{equation}
    det(D^{2}u)=\frac{E_{s}(x_{1},x_{2})}{E_{t}(\bigtriangledown{u}(x_{1},x_{2}))}, (x_{1},x_{2})\in\mathcal{X},
\end{equation}

\begin{equation}
    \bigtriangledown{u}(\partial{\mathcal{X}})=\partial{\mathcal{Y}},
\end{equation}

where $D^{2}u$ is the Hessian Matrix of the convex function $u$. The Eq. (2) is the boundary condition that represents the boundary points of source plane should be mapped to the boundary of target plane. Several numerical algorithms can be applied to solved this standard Monge-Amp\'{e}re equation such as Newton's method \cite{wu2014initial,froese2013convergent,feng2016freeform} and "time iteration method"\cite{feng2013designing,haker2004optimal,sulman2011efficient}. In this section, we employ a least-squares method proposed by C.R.PRINS et al.\cite{prins2015least} which is very efficient and robust for the optimal transport problem.

For the rectangular uniform irradiance distribution target, it's convenient to reverse the target and source plane by defining the target plane as the starting grid. After solving Eq. (1)(2), the equal energy source grid is obtained. As shown in Fig. 2(b), the source grid is divided into two parts for the refractive and reflective components and the target grid is redistributed into two unique grids for the mapping of two parts in the source plane respectively. Note that the two mapping presented in Fig. 2(b) are both curl-free, which can be simply proven. The emitted energy from the refractive part and reflective part are both mapped to the target plane for a prescribed distribution.

\subsection{Determine the entrance and TIR surfaces based on Genetic Algorithm (GA)}

In this section, we demonstrate the method to determine the analytic entrance surface and TIR surface. In our previous publication, the entrance surface is a simple sphere and the exit surface of TIR part is a plane which is zero degree of design freedom\cite{zhu2018catadioptric}. In this paper, we increased the design freedom by setting analytic entrance and TIR surface. Furthermore, the exit surface of TIR part is changed to a freeform surface for the intensity control of backside light.

Assuming that $f(x,y)$ denotes a specific analytic expression of the entrance surface, the aim is to find the optimal parameters of the analysis formula in order to make full usage of the existing design freedom. In this paper, the Genetic Algorithm (GA) is proposed to tackle the optimization problem. The merit function being optimized can be detailed as a collection of two components:

    \textbf{$\cdot$} \textit{Surface normal integrability condition}: Let $\textbf{I}$ labels the vector field of rays refracted from the entrance surface and $\textbf{O}$ denotes the vector field of emitted rays from freeform lens. For road illumination applications, the target plane is at a sufficient large distance from the freeform lens. The vector $\textbf{O}$ of exit rays can directly caculated from the target grid. According to  Karel Desnijder's work\cite{desnijder2019ray}, the integrability condition can be converted to:
    
    \begin{equation}
        (\bigtriangledown\times\textbf{N}')\cdot\textbf{I}=0,
    \end{equation}
    where \textbf{N}' is defined as \textbf{N}/(\textbf{N}$\cdot$\textbf{I}). By using the formal definition of the curl of s vector field, the Eq. (3) can be presented as a infinitesimal contour integral:
    
    \begin{equation}
        (\bigtriangledown\times\textbf{N}')\cdot\textbf{I}=\lim_{\bigtriangleup{S}\to0}\frac{1}{|\bigtriangleup{S}|}\oint_{\bigtriangleup\textbf{C}}\textbf{N}'\cdot{d\textbf{r}},
    \end{equation}
    where $\bigtriangleup{S}$ is an infinitesmal surface area with \textbf{I} as normal vector and $\bigtriangleup$\textbf{C} as a contour. Eq. (4) can be numerical represented so that the integrability condition can be quantified as a component of the merit function. The objective function of integrability condition can be represented as:
    
    \begin{equation}
    \begin{split}
        MF_{1}=\sum_{i,j}[\frac{\textbf{N}'_{i,j}+\textbf{N}'_{i+1,j}}{2}\cdot(\textbf{I}_{i+1,j}-\textbf{I}_{i,j}) + \frac{\textbf{N}'_{i+1,j}+\textbf{N}'_{i+1,j+1}}{2}\cdot(\textbf{I}_{i+1,j+1}-\textbf{I}_{i+1,j}) \\
        + \frac{\textbf{N}'_{i+1,j+1}+\textbf{N}'_{i,j+1}}{2}\cdot(\textbf{I}_{i,j+1}-\textbf{I}_{i+1,j+1}) +  \frac{\textbf{N}'_{i,j+1}+\textbf{N}'_{i,j}}{2}\cdot(\textbf{I}_{i,j+1}-\textbf{I}_{i,j})]^{2} .
        \end{split}
    \end{equation}
    
\begin{figure}[h!]
\centering\includegraphics[]{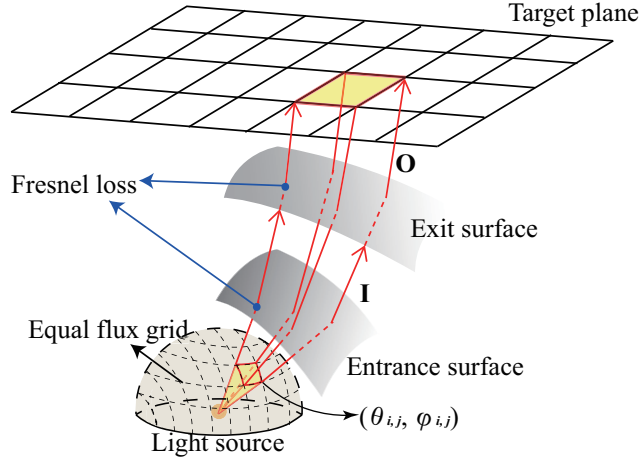}
\caption{Double surface freeform illumination system diagram.}
\end{figure}
 
   \textbf{$\cdot$} \textit{Transfer efficiency}: The transfer efficiency is defined as the ratio of flux at the target to the flux emitted from the source. The objective function component of transfer efficiency is aimed at the reduction of Fresnel losses and increase the efficiency. The conceptual diagram is presented in Fig. 3. For a incident natural light at medium interface, the energy transmissivity can be represented as:
\begin{equation}
       T=1-\frac{1}{2}[\frac{sin^{2}(\theta_{1}-\theta_{2})}{sin^{2}(\theta_{1}+\theta_{2})}+\frac{tan^{2}(\theta_{1}-\theta_{2})}{tan^{2}(\theta_{1}+\theta_{2})}],
    \end{equation}
where $\theta_{1}$, $\theta_{2}$ denote the incident angle and emergence angle respectively. The objective function of transfer efficiency is defined as the total transmissivity:

\begin{equation}
\begin{split}
       MF_{2}=\frac{\iint_{\mathcal{Y}}E_{t}(y_{1},y_{2})dy_{1} dy_{2}}{\iint_{\mathcal{X}}I(\theta,\phi)d\theta d\phi}
       =\frac{\sum_{i,j}I_{0}cos\theta_{i,j}\bigtriangleup{\Omega_{i,j}T_{1 i,j}T_{2 i,j}}}{\sum_{i,j}I_{0}cos\theta_{i,j}\bigtriangleup{\Omega_{i,j}}},
\end{split}
\end{equation}
where $T_{1}, T_{2}$ denote the transmissivity on the entrance and exit surface for each incident ray, $\bigtriangleup\Omega$ is the solid angle element of the mapping grid. 

The parameters of the entrance surface analysis formula can be determined based on Genetic Algorithm (GA) by using the merit function demonstrated above. For the TIR surface, the merit function contains only one components of surface normal integrability condition. It's worth noting that one must add a boundary condition that all backside rays incident angle on TIR surface should be larger than the total reflection angle. One can easily implement the optimization process on the Genetic Algorithm toolbox in Matlab. The specific design and optical prescription of entrance surface and TIR surface will be discussed later. 

\subsection{Construct the freeform surface}

In this section, we modify the freeform construction method proposed by Feng et al.\cite{feng2016freeform} to match the two surfaces problem. From the mapping demonstrated in the previous section, we can obtain the incident ray sequence and a target point sequence labeled as $\textbf{T}_{i,j}=(y_{1 i,j},y_{2 i,j},H)$. As presented in Fig. 4, \textbf{P} denotes the rays intersection point on the analytic entrance surface and \textbf{Q} denotes the freeform surface point. Assume that $\hat{\textbf{r}}_{i,j}$ represents the unit ray vector from \textbf{O} to $\textbf{P}_{i,j}$ and $\hat{\boldsymbol{\rho}}_{i,j}$ labels the unit ray vector from  $\textbf{P}_{i,j}$ to  $\textbf{Q}_{i,j}$. For far field illumination applications, the outgoing ray sequence can be caculated as $\hat{\textbf{O}}_{i,j}=\textbf{T}_{i,j}/|\textbf{T}_{i,j}|$. The normal vector field can be obtained from the Snell's law:

\begin{equation}
\begin{split}
       \hat{\textbf{N}}_{i,j}=\frac{\hat{\textbf{O}}_{i,j}-n\hat{\boldsymbol{\rho}}_{i,j}}{|\hat{\textbf{O}}_{i,j}-n\hat{\boldsymbol{\rho}}_{i,j}|},
\end{split}
\end{equation}
where $n$ denotes the lens material refractive index.

\begin{figure}[h!]
\centering\includegraphics[]{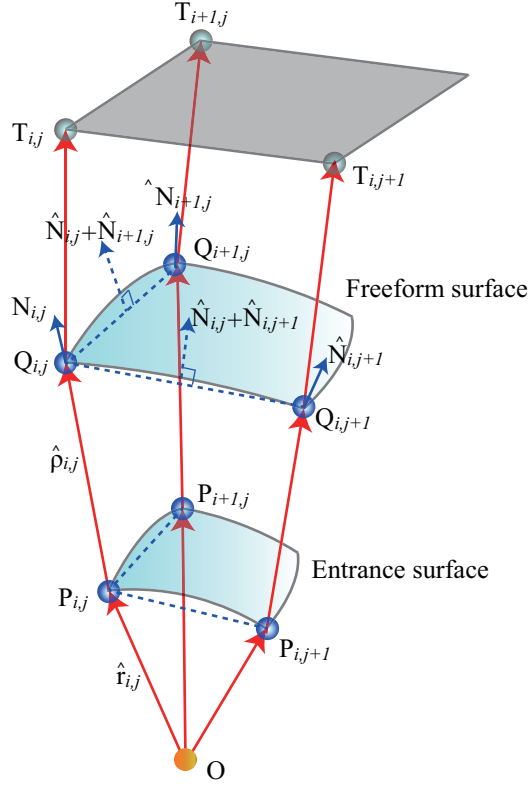}
\caption{Generating the freeform surface.}
\end{figure}

According to Feng's method\cite{feng2016freeform}, the freeform surface point and the surface normal are linked in a concise relationship:
\begin{equation}
\begin{split}
       (\textbf{Q}_{i+1,j}-\textbf{Q}_{i,j})\cdot(\hat{\textbf{N}}_{i+1,j}+\hat{\textbf{N}}_{i,j})=0,
\end{split}
\end{equation}
\begin{equation}
\begin{split}
       (\textbf{Q}_{i,j+1}-\textbf{Q}_{i,j})\cdot(\hat{\textbf{N}}_{i,j+1}+\hat{\textbf{N}}_{i,j})=0.
\end{split}
\end{equation}
The above equation means that the chord connecting the two adjacent points is perpendicular to the average of their normal vectors at these two points (see Fig. 4). The surface point $\textbf{Q}_{i,j}$ can be represented as $r_{i,j}\hat{\textbf{r}}_{i,j}+\rho_{i,j}\hat{\boldsymbol{\rho}}_{i,j}$, where $r_{i,j}$ and $\rho_{i,j}$ represent the distance from \textbf{O} to $\textbf{P}_{i,j}$ and the distance from $\textbf{P}_{i,j}$ to $\textbf{Q}_{i,j}$. The Eq. (9)(10) can be written as:

\begin{equation}
\begin{split}
       \hat{\boldsymbol{\rho}}_{i+1,j}\cdot(\hat{\textbf{N}}_{i+1,j}+\hat{\textbf{N}}_{i,j})\rho_{i+1,j}-\hat{\boldsymbol{\rho}}_{i,j}\cdot(\hat{\textbf{N}}_{i+1,j}+\hat{\textbf{N}}_{i,j})\rho_{i,j}=(\hat{\textbf{r}}_{i,j}r_{i,j}-\hat{\textbf{r}}_{i+1,j}r_{i+1,j})\cdot(\hat{\textbf{N}}_{i+1,j}+\hat{\textbf{N}}_{i,j}),
\end{split}
\end{equation}

\begin{equation}
\begin{split}
       \hat{\boldsymbol{\rho}}_{i,j+1}\cdot(\hat{\textbf{N}}_{i,j+1}+\hat{\textbf{N}}_{i,j})\rho_{i,j+1}-\hat{\boldsymbol{\rho}}_{i,j}\cdot(\hat{\textbf{N}}_{i,j+1}+\hat{\textbf{N}}_{i,j})\rho_{i,j}=(\hat{\textbf{r}}_{i,j}r_{i,j}-\hat{\textbf{r}}_{i,j+1}r_{i,j+1})\cdot(\hat{\textbf{N}}_{i,j+1}+\hat{\textbf{N}}_{i,j}).
\end{split}
\end{equation}
For all feature rays of the optimal transport mapping using the Eq. (11)(12), we can obtain a system of linear equations of $\hat{\boldsymbol{\rho}}$ which can be formulated as a single matrix equation:

\begin{equation}
\begin{split}
     \textbf{H}\boldsymbol{P}=\textbf{b},  
\end{split}
\end{equation}
where \textbf{H} is a sparse matrix containing all the coefficients of $\rho_{i,j}$. After setting one surface point to a given value in the colume vector \textbf{b}, we can obtain a least-squares solution from Eq. (13) as:

\begin{equation}
\begin{split}
     \boldsymbol{P}=(\textbf{H}^{T}\textbf{H})^{-1}\textbf{H}^{T}\textbf{b},  
\end{split}
\end{equation}
The algorithm is concretely introduced in Ref.\cite{feng2016freeform,herrmann1980least}. The freeform surface in the reflective part can be generated by using the same method demonstrated above.

\section{Design example and Result}

In this section, we design the multi-surface catadioptric freeform lens for off-axis road illumination. The design specification is the same with Ref.\cite{zhu2018catadioptric} for a 24m$\times$12m rectangular pattern. The parameters are shown in Table. 1. We use PMMA as the material of the lens due to its low cost and good manufacturability, and its refractive index is about 1.493. $X_{max}$ is the half-width of the rectangular illumination pattern in $x$ direction, $Y_{min}$ is the size of the rectangular illumination pattern for the part in sidewalk direction, and $Y_{max}$ is the size of the rectangular illumination pattern for street side. The acceptance angle is 90$^{\circ}$ (full angle of 180$^{\circ}$).

\begin{table*}
\begin{center}
\caption{Parameters of the road illumination system}
\setlength{\tabcolsep}{3mm}
\renewcommand\arraystretch{1}
\begin{tabular}{cccccc}
\hline
$X_{max}$(m) &  $Y_{max}$(m) & $Y_{min}$(m) & Acceptance angle X($^{\circ}$) & H(m) & Material\\
\hline
12 & 10 & 2 & 90 & 8 & PMMA\\
\hline
\end{tabular}
\end{center}
\end{table*}

We determine the entrance surface and TIR surface as elliptic paraboloid defined as $z=h-x^{2}/a^{2}_{1}-(y-b)^{2}/a^{2}_{2}$ for its multi-degree of design freedom and the simplicity of ray tracing. If one use surface with a complex analysis formula, the Newton's iteration or dichotomy method can be employed for the fast ray tracing. The designed catadioptric freeform lens is presented in Fig. 5.

\begin{figure}[h!]
\centering\includegraphics[]{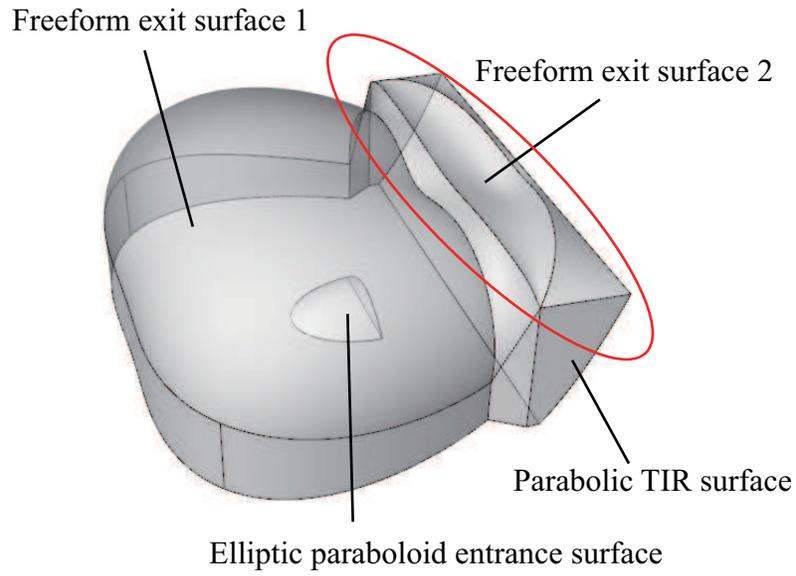}
\caption{Design result of the catadioptric freeform lens.}
\end{figure}

We align multiple luminaires along the rim of the road and check their optical performance. The comparison of the four surfaces design with the two surfaces design proposed in\cite{zhu2018catadioptric} is presented in Fig. 6. As shown in the irradiance distribution diagram, the four-surface design has a significant improvement on uniformity and the boundary quality. Here we use the relative standard deviation (RSD) and the transfer efficiency $\eta$ to evaluate the general performance for road illumination applications.The RSD is defined as:

\begin{equation}
\begin{split}
     RSD=\sqrt{\frac{1}{N_{p}}\sum_{i=1}^{N_{p}}[\frac{E_s(i)-E_0(i)}{E_0(i)}]^2},  
\end{split}
\end{equation}
where $N_p$ is the total number of sampling points inside effective analysis area on the target surface, $E_s(i)$ is simulation irradiance level for $i_{th}$ checking point on the target surface, and $E_0(i)$ is the desired illumination requirement of the checking point on the target plane. The transfer efficiency is defined as the ratio of flux at the target to the flux emitted from the source. In this paper, we defined the efficiency as the effective energy inside the illuminance pattern from -2m to 10m to the total energy emitted from light source. The performance comparison of the four-surface design and our proposed two-surface design is presented in Table. 2. The efficiency of the four-surface design result reached to 86.32\% with Fresnel losses considered.

\begin{figure}[h!]
\centering\includegraphics[width=14cm]{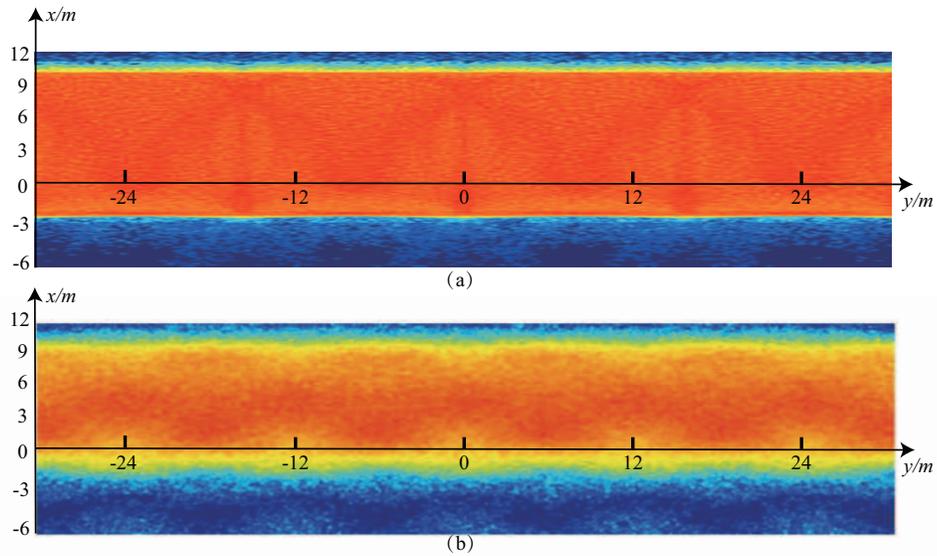}
\caption{(a) The irradiance distribution of the four surfaces design, (b) the irradiance distribution of the two surfaces design.}
\end{figure}

\begin{table*}
\begin{center}
\caption{Comparison of the two-surface design and the four-surface design.}
\setlength{\tabcolsep}{3mm}
\renewcommand\arraystretch{1}
\begin{tabular}{cccc}
\hline
  &  $RSD$ & $E_{ave}$(lx) & $\eta$ \\
\hline
Four-surface design & 0.14 & 29 & 86.32\%\\
\hline
Two-surface design & 0.19 & 29 & 81.81\%\\
\hline
\end{tabular}
\end{center}
\end{table*}

\section{Conclusion}
This paper presents an extension of the work proposed in\cite{zhu2018catadioptric}. We proposed a novel configuration of a four-surface catadioptric freeform lens for off-axis illumination applications. An efficient method combined with the ray mapping method, Genetic Algorithm and the surface reconstruction method is proposed to deal with the multi-surface design problem. The result shows a significant improvement on illuminance uniformity and transfer efficiency compared to the two-surface design. The presented design illustrate the feasibility and superiority of our proposed method for the ultra-efficient off-axis road illumination problem.

\section*{Funding}
Wuhan Municipal Science  and Technology Bureau (2017010201010110); Huazhong University of Science and Technology (2017KFYXJJ026); Natural Science Foundation of Hubei Province (2018CFB146); National Natural Science Foundation of China (61805088).

%%%%%%%%%%%%%%%%%%%%%%% References %%%%%%%%%%%%%%%%%%%%%%%%

%%%%%%%%%% If using BibTeX:
\bibliography{off-axis-road-illumination}

\end{document}